\documentclass[pra,12pt]{revtex4}
\usepackage{amssymb,amsfonts,amsmath,bm,graphics,epsfig}
\begin{document}

\title{Pre- and post-selected ensembles and time-symmetry in quantum mechanics}

\author{D. J. Miller}

\affiliation{Centre for Time, Department of Philosophy, University of Sydney NSW 2006, Australia and School of Physics, University of New South Wales, Sydney NSW 2052, Australia}

\email{D.Miller@unsw.edu.au}

\begin{abstract}
An expression is proposed for the quantum mechanical state of a pre- and post-selected ensemble, which is an ensemble determined by the final as well as the initial state of the quantum systems involved. It is shown that the probabilities calculated from the proposed state agree with previous expressions, for cases where they both apply. The same probabilities are found when they are calculated in the forward- or reverse-time directions. This work was prompted by several problems raised by Shimony recently in relation to the state, and time symmetry, of pre- and post-selected ensembles.  
\end{abstract}
\maketitle

\section{Introduction}

Quantum mechanics (QM) deals only with pre-selected ensembles, that is ensembles defined by a preparation outcome with a view to determining the probabilities of measurement outcomes in the future of the preparation event. Therefore QM is well-adapted to the intuitive concept of time, or more generally the A-theory of time, in which the past and the future have different status. The B-theory of time is an alternative, and counterintuitive, theory of time in which the past and the future have similar status. Modern physics is said by some (see Ref.~\cite{dickson,petkov} for discussions) to favour the ``block universe view" which is a B-theory of time. If the B-theory or block universe view is correct, it seems more natural to consider ensembles which are pre-selected by the outcome of a preparation experiment and also post-selected by the outcome of a subsequent measurement with a view to determining the probabilities of the results of a measurement (or sequence of measurements) at intervening times.

Interest in pre- and post-selected ensembles (PPSEs) began with the work of Aharonov, Bergmann and Lebovitz (ABL) \cite{abl} and has been pursued mainly by Aharonov and co-workers \cite{aharonovjphysa,avreview}. Sometimes QM for the PPSEs is referred to as the two-vector formalism \cite{aharonovjphysa} or time-symmetric quantum mechanics \cite{avreview}. It is possible that ideas prompted by PPSEs have wider ramifications for QM \cite{gruss,millerconf,aharonovgruss}.

Recently, Shimony has drawn two significant conclusions about PPSEs \cite{shimony1,shimony2,shimony3}. The first is that the ``generalised state" \cite{aharonovjphysa} said to be determined by the ``two vectors" specifying the initial and final conditions in a PPSE cannot be a regarded in the same way as the ``state" determined by the initial conditions for the pre-selected (only) ensemble in QM. The main reason \cite{shimony1,shimony2,shimony3} is that the probabilities for the outcomes of the intermediate measurements of observables in a PPSE can depend on the method of their measurement, whereas the ``state" in QM determines the probabilities of future measurement outcomes for any observable no matter how it is (faithfully) measured. In section 2, we propose a different way of defining a state for a PPSE which incorporates expressly the method which is used for the measurement of the intermediate observables. The proposal seems to go as far as one can in meeting the first of the two conclusions about PPSEs in Ref.~\cite{shimony1}.

The second conclusion concerns time-symmetry in relation to PPSEs. From the beginning, it had been thought \cite{abl} that, although QM had a built-in time asymmetry because it relied on initial conditions for subsequent properties, a PPSE could be regarded as time-symmetric for intermediate properties because it was selected by initial and final conditions. Two specific examples are given in Ref.~\cite{shimony1} which appear to show that the respective PPSE is not time-symmetric on either of two meanings of that term. Given that QM exhibits orthodox time-reversal symmetry \cite{wigner,sachs} and that a PPSE is defined in an apparently time-symmetric way, one would expect that a PPSE would preserve orthodox time-reversal symmetry. In Section III, it is argued that a PPSE does satisfy orthodox time-reversal symmetry. This is illustrated with reference to the two examples in Ref.~\cite{shimony1} in the Appendix. In Section IV, the new formalism developed in Section 2 is applied to the 3-box problem, a well-known example involving PPSEs.

\section{Definition of a ``state" for a PPSE}

Traditionally \cite{abl,aharonovjphysa,avreview}, a PPSE has been  defined only in terms of the Hilbert space of the quantum system that is being measured. In fact, the problem involves the tensor product of at least four Hilbert spaces: one for the quantum system that is being pre- and post-selected and measured, one for the apparatus involved in determining the pre-selection, one (or more) for the apparatuses involved in the measurement(s) of the intermediate observable(s) of the quantum system and one the for the apparatus involved in determining the post-selection. In the simplest possible case, the initial state of the system as a whole at some time $t_o$ can be specified by the state
\begin{equation}
|\Psi \rangle = |\psi \rangle \otimes |\alpha \rangle  \otimes
|\gamma \rangle  \otimes |\beta \rangle 
\end{equation}
where $|\psi \rangle $, $|\alpha \rangle $, $|\gamma \rangle $, and $|\beta \rangle $ are the (assumed non-degenerate) states of the quantum system, the pre-selection apparatus, the intermediate measurement apparatus (IMA) and the post-selection apparatus, respectively. In principle \cite{shimony3} it is necessary to deal with degeneracies and mixtures in relation to the initial and final states but the resulting expressions are not simple \cite{abl} and do not add anything new of a conceptual nature. The simple case of a non-degenerate initial state assumed here, along with similar assumptions at subsequent stages, are sufficient to deal with all the cases in the relevant literature.

Beginning with the above state, a practical means of carrying out the pre-selection at time $t_a > t_o$ could involve a measurement for observable $A$ of the quantum system with eigenvalues $|a_j \rangle $ in which it was possible to filter out of all but one of the possible measurement outcomes. The latter step can be represented by the projection onto $\hat{P}_{a_i} \otimes \hat{P}_{\alpha_{i}} = |a_i \rangle \langle a_i | \otimes |\alpha_{i} \rangle \langle \alpha_{i} |  $ (assuming $|a_i \rangle $ and $|\alpha_i \rangle $ are non-degenerate). Thus the pre-selection can be characterised by the evolution
\begin{eqnarray}
 |\psi \rangle \otimes |\alpha \rangle & \rightarrow  \sum_{j} \langle a_j |\psi \rangle | a_j \rangle \otimes |\alpha_j \rangle \;\; \text{(measurement interaction)}\\
& \rightarrow  | a_i \rangle \otimes |\alpha_i \rangle \;\; \text{(projection)}
\end{eqnarray}
and the pre-selected state at time $t_a$ is then
\begin{equation}
|A;t_a \rangle = |a_i \rangle  \otimes |\alpha_i \rangle  \otimes
|\gamma \rangle  \otimes |\beta \rangle  .
\end{equation}

We take the intermediate measurement to be performed during the time $t_c - \epsilon $ and $t_c$, with $t_c > t_a$, $\epsilon > 0$, and to be of observable $C$ with possibly degenerate eigenvalues $c_k$ and corresponding eigenstates $| c_{kl} \rangle $, where the $| c_{kl} \rangle $ for $l \in \{\ldots\}_k $ span the subspace with eigenvalue $c_k$. In order to deal with all the cases of Ref.~\cite{shimony1}, it is necessary to consider a rather general form of the intermediate measurement. The measurement can be thought of as taking place in two steps: firstly, for each eigenvalue $c_k$, there is a preferred set of basis states $| c_{kl} \rangle$, each of which is independently transformed into a new state within the same eigenspace corresponding to $c_k$. The states of the IMA record, in the original preferred basis, both the components $| c_{kl} \rangle$ of the original state and the components $| c_{km} \rangle$ of the new state. Thus the interaction with the IMA for the measurement of $C$, including evolution from $t_a$ to $t_c$, leads to the initial state for those two systems at time $t_c$  
\begin{equation}
\hat{U}(t_c,t_a) |a_i \rangle \otimes |\gamma \rangle  =  \sum_{k,l} \langle c_{kl}|\hat{U}(t_c,t_a) |a_i \rangle \sum_m d^{k}_{lm} | c_{km} \rangle | \otimes |\gamma^{k}_{lm} \rangle
\end{equation}
where $\hat{U}(t_c,t_a)$ is the time evolution operator for the quantum system and IMA, the $|\gamma^{k}_{lm} \rangle $ for $l,m \in \{\ldots\}_k $ are all states of the measuring apparatus registering the experimental outcome with eigenvalue $c_k$ and $\sum_m |d^{k}_{lm}|^2 = 1$. For an eigenspace which is doubly-degenerate in spin, for example, one could imagine first passing the quantum system in state $|a_i \rangle$ through a Stern-Gerlach apparatus and then passing each resulting beam through another Stern-Gerlach apparatus that is oriented at an independently chosen angle to determine the $d^{k}_{lm}$. The resulting paths then correspond to the $|\gamma^{k}_{lm} \rangle $. Some specific cases are considered in the following subsections.

Finally, the post-selection at time $t_b > t_c$ involves the measurement of observable $B$ for the quantum system with (non-degenerate) eigenvalues $|b_i \rangle $ and the filtering out of all but one of the possible measurement outcomes, which can be represented by the projection onto $\hat{P}_{b_{j}} \otimes \hat{P}_{\beta_{j}} = |b_j \rangle \langle b_j | \otimes |\beta_{j} \rangle \langle \beta_{j} |  $. The final measurement occupies the period $t_b - \eta > t_c$ to $t_b$, $\eta >0$. Thus the post-selection can be characterised by the evolution of each of the states in the summation in Eq.~(5) with the state $| \beta\rangle $ of the final measuring apparatus, as follows
\begin{subequations}
\begin{eqnarray}
 \hat{U}(t_b,t_c) | c_{km} \rangle \otimes |\beta \rangle & \rightarrow  \sum_{i} \langle b_i |\hat{U}(t_b,t_c)|c_{km} \rangle | b_i \rangle \otimes |\beta_i \rangle \;\; \text{(measurement interaction)}\\
& \rightarrow  | b_j \rangle \otimes |\beta_j \rangle \;\; \text{(projection)}.
\end{eqnarray}
\end{subequations}

Combining Eqs.~(5) and (6), the final state at time $t = t_b$ is
\begin{equation}
| B;t_b \rangle = \sum_{k,l} \langle c_{kl}|\hat{U}(t_c,t_a) |a_i \rangle \sum_m d^{k}_{lm} \langle b_j|\hat{U}(t_b,t_c) |c_{km} \rangle | b_j \rangle \otimes |\alpha_i \rangle \otimes |\gamma^{k}_{lm} \rangle \otimes |\beta_j \rangle .
\end{equation}

As emphasised in Refs.~\cite{shimony1,shimony2,shimony3}, the probabilities for the outcomes of the intermediate measurement depend on the details of the measurement performed on the quantum system by the IMA. The intermediate measurement can always be characterised by a set $\Gamma =\{\hat{\Gamma}_i\}$ of positive operators which constitute a resolution of the identity operator. In the present case $\Gamma =  \{|\gamma^{k}_{lm} \rangle \langle \gamma^{k}_{lm} | \}$. It follows that if it is possible to specify a ``state" for a PPSE which applies in general, the state must depend on $\Gamma$ and since it must provide probabilities for the intermediate measurement outcomes, it must be linear in the $\{\hat{\Gamma}_i\}$. Therefore we propose the following density operator to represent the state of the PPSE in the general case
\begin{equation}
\hat{\rho}(A,\Gamma ,B) = \sum_{i} D_{i}(A,\Gamma ,B) 
\hat{\Gamma}_i
\end{equation}
where
\begin{subequations}
\begin{eqnarray}
 D_{i}(A,\Gamma ,B) & = & \frac{ | \langle B;t_c| \hat{\Gamma}_i |A;t_c \rangle |^2}
{\sum_{i}| \langle B;t_c|\hat{\Gamma}_i  |A;t_c \rangle |^2} \\
 & = & \frac{| \langle B;t_b|\hat{U}(t_b,t_c) \hat{\Gamma}_i \hat{U}(t_c,t_a) |A ;t_a\rangle |^2}
{\sum_{i}| \langle B;t_b|\hat{U}(t_b,t_c) \hat{\Gamma}_i \hat{U}(t_c,t_a) |A ;t_a\rangle |^2} .
\end{eqnarray}
\end{subequations}

Once the density operator has been defined, the probability for the outcome $\Gamma_j$ of the intermediate measurement is given by the usual expression
\begin{subequations}
\begin{eqnarray}
\text{Prob}[\Gamma_j|\hat{\rho}(A,\{\hat{\Gamma_i}\} ,B)] & = & \text{Tr}  (\hat{\rho}(A,\{\hat{\Gamma_i}\} ,B) \hat{\Gamma_j})  \\
& = &  \frac{| \langle B|\hat{U}(t_b,t_c) \hat{\Gamma_j} \hat{U}(t_c,t_a) |A \rangle |^2 }{\sum_{i} | \langle B|\hat{U}(t_b,t_c) \hat{\Gamma_i} \hat{U}(t_c,t_a)|A \rangle |^2 } .
\end{eqnarray}
\end{subequations}
  
It is suggested that Eq.~(8) is the preferable way of specifying the state of a PPSE. Two things are worth noting. Firstly, Eq.~(8) is a contextual probability measure. A contextual probability measure is unacceptable for the calculation of properties which depend only on the preparation of a quantum system \cite{peres}. For the PPSE, the calculation of properties other than $\{\hat{\Gamma}_i\}$ is irrelevant because the IMA is fixed for the PPSE and the stricture against a contextual probability measure does not apply. Secondly, Eq.~(8) leads to probabilities for various intermediate measurements which manifestly obey the laws of probability \cite{hughes} for the set of events confined to the outcomes for the particular choice of IMA.  Previous expressions \cite{aharonovjphysa,avreview} for the probabilities of intermediate measurements for PPSEs have not been expressed in terms of a density operator and have had to be written in different ways for different forms of the IMA \cite{shimony2}. Those different forms follow naturally from the general expression given in Eqs.~(8) and (9) and are in agreement with other work which we now show expressly by reference to the cases considered in Ref.~\cite{shimony2}.

\subsection{Non-degenerate case}

In this case, the IMA registers only the eigenvalues which means the only states of the IMA are $| \gamma^{n}_{nn} \rangle $ which we can write as $| \gamma^{n} \rangle $. The positive operator $\hat{\Gamma}_n = | \gamma^{n} \rangle \langle \gamma^{n} |$ for the IMA corresponds to the eigenvalue $c_n$ for the quantum system and $\Gamma = \{| \gamma^{n} \rangle \langle \gamma^{n} |\}$. Also there is, for each eigenvalue $c_k$, only one eigenstate $| c_{k1} \rangle$ ($\{\ldots\}_k$ has only one member) which we we can write as $| c_{k} \rangle $ and $d^{k}_{lm} = \delta_{kl}\delta_{lm}$ for all $k$.  The probability that the IMA registers eigenvalue $c_n$ becomes
\begin{subequations}
\begin{eqnarray}
\text{Prob}[c_n|\hat{\rho}(A,\Gamma ,B)] & = & \text{Tr}  (\hat{\rho}(A,\Gamma ,B) |\gamma^{n} \rangle \langle \gamma^{n} |)  \\
& = &  \frac{| \langle b_j|\hat{U}(t_b,t_c) |c_{n} \rangle |^2 | \langle c_{n}|\hat{U}(t_c,t_a) |a_i \rangle |^2 }{\sum_{k} | \langle b_j|\hat{U}(t_b,t_c) |c_{n} \rangle |^2 | \langle c_{n} |\hat{U}(t_c,t_a)|a_i \rangle |^2 } 
\end{eqnarray}
\end{subequations}
in agreement with Eq.~(14) of Ref.~\cite{shimony2}.

\subsection{Degenerate cases}

In this case, $\{\ldots\}_k$ has more than one member. Firstly there is the case when the IMA does not distinguish between the eigenvectors which span the eigenspace and therefore still only registers $\gamma^{k}$ for each eigenvalue $c_k$ so that $d^{k}_{lm} = \delta_{kl}\delta_{lm}$ for all $k$ and $\Gamma = \{| \gamma^{n} \rangle \langle \gamma^{n} |\}$, as before. From Eq.~(10) the probability that the IMA registers eigenvalue $c_n$ is
\begin{subequations}
\begin{eqnarray}
\text{Prob}[c_n|\hat{\rho}(A,\Gamma ,B)] & = & \text{Tr} (\hat{\rho}(A,\{\gamma^{k}\} ,B) |\gamma^{n} \rangle \langle \gamma^{n} |)  \\
& = &  \frac{| \sum_{l} \langle b_j|\hat{U}(t_b,t_c) |c_{nl} \rangle \langle c_{nl}|\hat{U}(t_c,t_a) |a_i \rangle |^2 }{ \sum_{k} | \sum_{l} \langle b_j|\hat{U}(t_b,t_c) |c_{kl} \rangle \langle c_{kl}|\hat{U}(t_c,t_a) |a_i \rangle |^2} .
\end{eqnarray}
\end{subequations}
in agreement with Eq.~(19) of Ref.~\cite{shimony2}.

Secondly there is the case when the the IMA distinguishes between the eigenvectors which span the eigenspace and therefore registers different $\gamma^{k}_{l}$ for $l \in \{\ldots\}_k $ but $d^{k}_{lm} = \delta_{lm}$ for all $k$ and $\Gamma = \{| \gamma^{k}_{l} \rangle \langle \gamma^{k}_{l} |\}$. Then the probability that the intermediate measurement registers eigenvalue $c_n$ is
\begin{subequations}
\begin{eqnarray}
\text{Prob}[c_n|\hat{\rho}(A,\Gamma ,B)] & = & \sum_l \text{Tr} (\hat{\rho}(A,\Gamma ,B) | \gamma^{n}_{l} \rangle \langle \gamma^{n}_{l} |)  \\
& = &  \frac{\sum_l | \langle b_j|\hat{U}(t_b,t_c) |c_{nl} \rangle \langle c_{nl}|\hat{U}(t_bc,t_a) |a_i \rangle |^2 }{ \sum_{k} \sum_{l} | \langle b_j|\hat{U}(t_b,t_c) |c_{kl} \rangle \langle c_{kl}|\hat{U}(t_c,t_a) |a_i \rangle |^2} .
\end{eqnarray}
\end{subequations}
in agreement with Eq.~(22) of Ref.~\cite{shimony2}.

Finally there is the case when the the IMA distinguishes between the original and final basis vectors in the eigenspace and therefore registers $\gamma^k_{lm}$ for the eigenvalue $c_k$ and $\Gamma = \{| \gamma^{k}_{lm} \rangle \langle \gamma^{k}_{lm} |\}$. Then the probability that the intermediate measurement registers eigenvalue $c_n$ is
\begin{subequations}
\begin{eqnarray}
\text{Prob}[c_n|\hat{\rho}(A,\Gamma ,B)] & = & \sum_{r,s} \text{Tr} (\hat{\rho}(A,\Gamma ,B) | \gamma^n_{rs} \rangle \langle \gamma^n_{rs} |)  \\
& = &  \frac{\sum_{r,s} | \langle b_j|\hat{U}(t_b,t_c) |c_{ns} \rangle d^{n}_{rs} \langle c_{ns}|\hat{U}(t_c,t_a) |a_i \rangle |^2 }{ \sum_{k}  \sum_{r,s} | \langle b_j|\hat{U}(t_b,t_c) |c_{kl} \rangle d^{n}_{rs} \langle c_{kl}|\hat{U}(t_c,t_a) |a_i \rangle |^2} .
\end{eqnarray}
\end{subequations}
This case is not considered in Ref.~\cite{shimony2} but it is used implicitly in Ref.~\cite{shimony1}.

The main conclusion of this section is that the expression in Eq.~(8) satisfies the requirements for the ``state" of a PPSE which is pre-selected by the projection given in Eq.~(4) and post-selected by the projection given in Eq.~(7) with the intermediate measurement specified by $\{\gamma^{k}_{lm}\}$. The PPSE state defined in that way satisfies one of the main conclusions of Refs.~\cite{shimony1,shimony2,shimony3}, namely that the state of a PPSE must depend on the nature of the intermediate measurement and cannot be characterised by the pre- and post-selections alone. In Section IV, it will be argued that the above formalism clarifies the main issues in a much-discussed example for PPSEs, the 3-box problem. Before doing so, the questions about time-symmetry raised in Refs.~\cite{shimony1,shimony2,shimony3} will be considered.

\section{Time-symmetry and PPSEs}

Two of the conclusions of Ref.~\cite{abl} were that (i) the laws of probability for PPSE's are time-symmetric and (ii) the laws for retrodiction are the same as the laws for prediction \cite{abldetail}. Ref.~\cite{abl} deals only with the case when both the quantum states and the observables remain constant in time. In Refs.~\cite{shimony1,shimony2,shimony3} less restrictive conditions are considered and different conclusions are reached. Firstly, it is necessary to be clear about the meaning of time-symmetry that is being used.

There is not universal consensus on the meaning of time-symmetry either in classical physics or in quantum physics (for recent discussions and references to the literature, see Ref.~\cite{malm} and Ref.~\cite{holster} respectively). The orthodox position is that a system is time-symmetric if it satisfies motion reversal invariance \cite{wigner,sachs} and that is the meaning which will be adopted in the following. Cocke \cite{cocke} has dealt with time-symmetry in relation to the ABL formalism for the non-degenerate case in terms of orthodox time-reversal symmetry and the following agrees with his results for that case. 

The starting point for orthodox time-reversal symmetry is that the symmetry applies if the Hamiltonian satisfies $\hat{H} = \hat{\Theta } \hat{H} \hat{\Theta }^{-1}$, where $\hat{\Theta } $ is the time-reversal operator which produces the time-reversed state $| \tilde{X} \rangle $ of any state $| X \rangle $: $| \tilde{X} \rangle = \hat{\Theta }| X \rangle $. It follows from the form of the time-evolution operator that a state is left unchanged by propagation forwards in time, followed by time reversal of the resulting state, the same propagation forwards in time again and a restoration of the original time sense \cite{wigner}. That is, 
\begin{equation}
\hat{\Theta }^{-1}\hat{U}(\Delta t) \hat{\Theta }\hat{U}(\Delta t) = \hat{I} \text{    or     } \hat{\Theta }^{-1}\hat{U}(\Delta t) \hat{\Theta } = \hat{U}^{\dag}(\Delta t) = \hat{U}(- \Delta t) 
\end{equation}
where $\hat{I}$ is the identity operator in the relevant Hilbert space.

In principle, there are eight ways of calculating the probabilities for a PPSE selected by states $|A \rangle $ and $|B \rangle $:
\begin{eqnarray*}
 & (i)  \;\;  |A \rangle \stackrel{\hat{U}(t_b,t_a)}{\rightarrow} |B \rangle \;\; 
(ii) \;\; |A \rangle \stackrel{\hat{U}^{\dag}(t_b,t_a)}{\leftarrow} |B \rangle \;\;
(iii) \;\; |\tilde{B} \rangle \stackrel{\hat{U}(t_b,t_a)}{\rightarrow} |\tilde{A} \rangle  \;\;  
(iv) \;\; |\tilde{B} \rangle \stackrel{\hat{U}^{\dag}(t_b,t_a)}{\leftarrow} |\tilde{A} \rangle  \\ 
 & (v)  \;\;  |\tilde{A} \rangle \stackrel{\hat{U}(t_b,t_a)}{\rightarrow} |\tilde{B} \rangle \;\; (vi) \;\; |B \rangle \stackrel{\hat{U}(t_b,t_a)}{\rightarrow} |A \rangle \;\; 
(vii) \;\; |\tilde{A} \rangle \stackrel{\hat{U}^{\dag}(t_b,t_a)}{\leftarrow} |\tilde{B} \rangle \;\; (viii) \;\; |B \rangle \stackrel{\hat{U}^{\dag}(t_b,t_a)}{\leftarrow} |A \rangle 
\end{eqnarray*}
where time is increasing from the left to right of each pair and the arrow indicates the direction of time evolution of the states. If orthodox time-reversal symmetry is satisfied the four entries in each row are equivalent but the two rows are equivalent only if the states are left invariant under $\hat{\Theta}$.

Since orthodox time-reversal symmetry requires that process (i) above is equivalent to either process (ii) or (iii), the time reversal of the process of selecting a PPSE by process (i) above can be calculated by considering either process (ii) or (iii), whichever is more convenient. For the first alternative, the same density operator for the PPSE should be obtained from Eq.~(8) using either the conventional process (starting from $|A:t_a \rangle$ and calculating $| B;t_b \rangle $ using process (i) above) or starting from $| B;t_b \rangle $ and calculating $|A:t_a \rangle$ using process (ii) above which involves beginning with the state $| B;t_b \rangle $ at time $t_b$, projecting it using $\hat{P}_{b_{j}} \otimes \hat{P}_{\beta_{j}}$ (which leaves $| B \rangle $ unchanged), propagating $| B \rangle $ in the reverse time direction using $\hat{U}^{\dag}(t_b,t_a) = \hat{U}(t_a,t_b)$ ($t_b > t_a$) and projecting it by $\hat{P}_{a_i} \otimes \hat{P}_{\alpha_{i}}$ to obtain $|A:t_a \rangle$. For the second alternative, the same density operator of the PPSE should should be obtained from Eq.~(8) using either the conventional process or starting from $|\tilde{ B} \rangle = \hat{\Theta} | B \rangle $ and calculating $|A:t_a \rangle$ using process (iii) above which involves beginning with the time-reversed state $|\tilde{ B} \rangle = \hat{\Theta} | B \rangle $ at time $t_b$, projecting it using the time-reversal of $\hat{P}_{b_{j}} \otimes \hat{P}_{\beta_{j}}$ (which leaves $| \tilde{ B} \rangle $ unchanged), propagating $| \tilde{ B} \rangle $ in the forward-time direction using the same $\hat{U}(t_b,t_a)$ (note that $\hat{U}$ is time-translation invariant) and projecting it by the time-reversal of $\hat{P}_{a_i} \otimes \hat{P}_{\alpha_{i}}$ and finally time-reversing the result to obtain $|A:t_a \rangle$. 

At this point, we refer to the alternative definitions of time symmetry for a PPSE considered in Ref.~\cite{shimony1} which, it should be remembered, was directed towards considering, and ultimately rejecting, the idea that a PPSE can be characterised by the pair of vectors in the Hilbert space of the quantum system alone. For that reason perhaps both of the definitions of time symmetry differ from those given above. They both rely \cite{shimony1} on the idea that, for the purposes of considering the reverse-time case, the state of the IMA at time $t_b$ should be {\it reset} to the state $| \gamma \rangle $ that it was in at time $t_a$ for the purposes of considering the forward-time case. That is instead of state $| B \rangle $ in Eq.~(7) it is said in Ref.~\cite{shimony1} above one should use the state
\begin{equation}
| B' \rangle = | b_j \rangle \otimes | \gamma \rangle
\end{equation}
This would be an attractive proposal if the PPSE could be characterised in just the Hilbert space of the quantum system alone but not if the PPSE depends on the details of the IMA which is the ultimate conclusion of Ref.~\cite{shimony1} and which is incorporated in the expression for the state of the PPSE proposed above. Therefore the approach to time-reversal in Ref.~\cite{shimony1} is not to be preferred over the conventional approach.

It is fairly clear that if the conventional approach to time-reversal is adopted the definition of PPSE is time-symmetric as originally stated \cite{abl,cocke}. Nevertheless, pre- and post-selections do involve non-unitary projections of states and so it seems worthwhile to show that both of the counterexamples from Ref.~\cite{shimony1} are indeed time-symmetric on the conventional view. This is done in Appendices A and B.

It is worth noting that it is said \cite{shimony4} that the propagation backward through the interval of the intermediate measurement (as defined here) $t_c$ to $t_c - \epsilon$ ``is illegitimate from the standpoint of standard quantum mechanics, since the system interacts with the measuring apparatus" during this interval. The same comment could be made for propagation backward through the time interval of the final measurement  $t_b$ to $t_c - \eta $. However the interaction between the measuring apparatus and the quantum system is unitary; it is the second step of the measurement, which projects the quantum system plus apparatus onto the observed state which is non-unitary.  Therefore, provided the Hamiltonian describing the interaction is time symmetric, i.e. $\hat{H} = \hat{\Theta } \hat{H} \hat{\Theta }^{-1}$, there is no reason why the interaction step involved in the IMA cannot be considered in the reverse direction of time. Of course, the original state (used in the forward-time propagation) will not be restored by the backward evolution of the projected state. Nevertheless, as discussed above and in the Appendices, conventional time-symmetry ensures that the same PPSE results from consideration in both time directions. 

\section{Application to the 3-box problem}

From the beginning, the study of PPSEs have led to counterintuitive results which have provoked discussion, for example see \cite{aaaprl} and \cite{bub}. The aim of the present section is to show that the definition of a state for a PPSE given in Eq.~(8) avoids some of the counterintuitive results. This is done by reference to a specific example, known as the 3-box problem \cite{aharonovjphysa,aaaprl}. For recent work and a partial list of references to earlier work on the 3-box problem, see \cite{ravon,fink,millerfof}.

In the 3-box problem, one considers three non-degenerate states $| X \rangle $, $| Y \rangle $ and $| Z \rangle $, which can be imagined to be ``boxes" in which a quantum system can be found. The pre-selection state of the quantum system is
\begin{equation}
| A \rangle = \frac{1}{\sqrt{3}} (| X \rangle + | Y \rangle + | Z \rangle) 
\end{equation}
and the post-selection state is
\begin{equation}
| B \rangle = \frac{1}{\sqrt{3}} (| X \rangle + | Y \rangle - | Z \rangle) .
\end{equation}
The intermediate measurement involves determining whether, for either $P$ = $X$, $Y$ or $Z$, the quantum system is in state $| P \rangle $, often described as ``looking" in box $P$. If the state of the IMA when the quantum system is found in box $P$ is $| \gamma^P \rangle $, the set $\Gamma $ describing the IMA consists of $| \gamma^P \rangle \langle \gamma^P |$ and $\hat{I} - | \gamma^P \rangle \langle \gamma^P | $, where $\hat{I}$ is the identity operator for the Hilbert space of the intermediate measurement apparatus. On the present approach the state of the PPSE in this case is, from Eq.~(8),
\begin{eqnarray}
 \hat{\rho}(A,\Gamma ,B) & = & | \langle b|\hat{U}(t_b,t_c) |X \rangle  \langle X|\hat{U}(t_c,t_a) |a \rangle |^2 | \gamma^X \rangle \langle \gamma^X |  \nonumber \\
& &  +
| \langle b|\hat{U}(t_b,t_c) (\hat{I} -| X \rangle )(\hat{I} - \langle X |) \hat{U}(t_c,t_a) |a \rangle |^2 (\hat{I} - | \gamma^X \rangle \langle \gamma^X |).
\end{eqnarray}
Applying Eq.~(19) with Eq.~(10), one finds the usual values \cite{aaaprl,aharonovjphysa} for the probabilities, which are given in Table 1.
\begin{table}
\caption{Probabilities of finding the quantum system when looking in either box $A$, $B$ or $C$ for the PPSE involved in the 3-box problem}
\vspace{5mm}
\begin{tabular}{|c|c|c|c|} \hline 
$P$ & Prob$_P(X)$ & Prob$_P(Y)$ & Prob$_P(Z)$ \\ \hline 
$X$ & 1  & 0  & 0 \\ \hline 
$Y$ & 0  & 1  & 0 \\ \hline 
$Z$ & $\frac{1}{3} $ & $\frac{1}{3} $ & $\frac{1}{3}$ \\ \hline  
\end{tabular}

\end{table}

It has been argued \cite{vaidfof} that because of the results in the first two rows of Table 1, the probability of finding the quantum system in box $A$ is unity {\it and} the the probability of finding the quantum system in box $B$ is unity which is contrary to the normal restriction that the sum of the probabilities of all the alternatives must be unity. That result does not follow from Table 1 on the present view because the results of each row are for three different ensembles with different states given by Eq.~(19) with $P = X$, $P = Y$ or $P = Z$ respectively. Thus the problems of the 3-boxes are avoided by adopting the version of one of the conclusions of Ref.~\cite{shimony1,shimony2,shimony3} that is set out above, namely, that the state of a PPSE depends on the details of the IMA.

\section{Conclusion}

The present work is in agreement with one of the main conclusions of Refs.~\cite{shimony1,shimony2,shimony3}, namely that the initial and final states of the quantum system that is measured are not sufficent to characterise a PPSE because a PPSE is crucially dependent on the nature of the intermediate measuring process(es). We have suggested a formal expression, given in Eqs.~(8) and (9), for the state of a PPSE which depends expressly on the nature of the IMA. That definition of a state for a PPSE avoids some of the counterintuitive conclusions that have been made about PPSEs in the past.

The second conclusion relates to time symmetry and PPSEs. Whether or not time symmetry applies to PPSE's hinges of course on what is meant by the term ``time symmetry". In Ref.~\cite{shimony1}, Shimony has considered two possible meanings of that term for the particular case of a PPSE. The two meanings seem to be motivated by the claim that was being assessed in Ref.~\cite{shimony1}, namely that the initial and final states of the quantum system that is measured are sufficient to characterise a PPSE, in which case either meaning is plausible and physically appealing. The conclusion was that a PPSE fails to satisfy time symmetry on either account.

Once one rejects the idea that the initial and final states of the quantum system are sufficent to characterise a PPSE, there is no reason not to adopt the conventional meaning that ``time symmetry" is equivalent to motion-reversal invariance. We have confirmed that PPSE's satisfy time symmetry for the counter-examples of Ref.~\cite{shimony1} if one adopts that definition.

\appendix
\section{Consideration of the first counter-example of Ref.~\cite{shimony1}}

In our notation, the example involves the intermediate measurement of observable $C$ which has a non-degenerate eigenvalue $c_0$ and corresponding eigenstate $|c_{00} \rangle $ and a doubly-degenerate eigenvalue $c_1$ with the corresponding eigenspace spanned by $|c_{11} \rangle $ and $|c_{12} \rangle $.  Since the states of the apparatus involved in the pre- and post-selection are non-degenerate, explicit reference to them can be dropped in the following for convenience. The pre-selection is implemented by projection of the quantum system at time $t_a$ leading to the following initial state for the quantum system and IMA 
\begin{equation}
| A \rangle = | a_i ) \otimes | \gamma \rangle = \frac{1}{\sqrt{2}} (| c_{00} \rangle + | c_{11} \rangle ) \otimes | \gamma \rangle.
\end{equation}

The states of the quantum system and the intermediate measuring apparatus $M_2$ are are assumed to be unchanged in time, apart from a phase factor which can be ignored, except during the interaction between the quantum system and the IMA during the time, in our case, $t_c - \epsilon > t_a$ to $t_c  < t_b$, $\epsilon > 0$. The Hamiltonian given in Ref.~\cite{shimony1} describing the interaction between the quantum system and the IMA has eigenvalues $E =0$ and $E = \pm g$ where $g$ measures the strength of the interaction. The corresponding energy eigenstates are
\begin{subequations}
\begin{eqnarray}
|\tau^{1}_{1} \rangle & = & -(d^{1}_{12})^* |c_{11} \rangle \otimes |\gamma^{1}_{11} \rangle + (d^{1}_{11})^* |c_{12} \rangle \otimes |\gamma^{1}_{12} \rangle)  \\
|\tau^{1}_{2} \rangle & = & (d^{1}_{22})^* |c_{11} \rangle \otimes |\gamma^{1}_{21} \rangle - (d^{1}_{21})^* |c_{12} \rangle \otimes |\gamma^{1}_{22} \rangle)    
\end{eqnarray}
\end{subequations}
for $E=0$ and
\begin{subequations}
\begin{eqnarray}
|\sigma^{0}_{\pm} \rangle & = & \frac{1}{\sqrt{2}}[|c_{00} \rangle \otimes |\gamma \rangle \pm |c_{00} \rangle \otimes |\gamma^{0}_{00} \rangle ]  \\
|\sigma^{11}_{\pm} \rangle & = & \frac{1}{2}[|c_{11} \rangle \otimes |\gamma \rangle \pm (d^{1}_{11} |c_{11} \rangle \otimes |\gamma^{1}_{11} \rangle + d^{1}_{12} |c_{12} \rangle \otimes |\gamma^{1}_{12} \rangle) ]  \\
|\sigma^{12}_{\pm} \rangle & = & \frac{1}{\sqrt{2}}[|c_{12} \rangle \otimes |\gamma \rangle \pm (d^{1}_{21} |c_{11} \rangle \otimes |\gamma^{1}_{21} \rangle + d^{1}_{22} |c_{12} \rangle \otimes |\gamma^{1}_{22} \rangle) ]  
\end{eqnarray}
\end{subequations}
for $E=\pm g$.

Expressing the initial state in terms of the energy eigenstates, it follows that if $\epsilon = \pi/2g$, the interaction between the quantum system and the IMS causes the evolution (in the forward-time direction)
\begin{equation}
\hat{U}(t_c, t_c - \epsilon)| A \rangle = -\frac{i}{\sqrt{2}} [| c_{00} \rangle \otimes | \gamma^{0}_{00} \rangle + d^{1}_{11} |c_{11} \rangle \otimes |\gamma^{1}_{11} \rangle + d^{1}_{12} |c_{12} \rangle \otimes |\gamma^{1}_{12} \rangle ].
\end{equation} 

The post-selection is implemented at time $t_b > t_c$ by projection of the quantum system onto the state
\begin{equation}
| b \rangle = \frac{1}{\sqrt{2}} (| c_{00} \rangle + | c_{12} \rangle)
\end{equation}
so that from Eq.~(7) after the post-selection projection, the state of the quantum system and IMA is
\begin{equation}
| B \rangle  = - \frac{i}{\sqrt{N}} ( | c_{00} \rangle \otimes | \gamma^{0}_{00} \rangle + d^{1}_{11} | c_{12} \rangle \otimes | \gamma^{1}_{11} \rangle ) 
\end{equation}
where $N = 1 + |d^{1}_{11}|^2$. In accordance with Section II above, the state of the PPSE is the density operator
\begin{equation}
\hat{\rho}(A,\Gamma ,B)  = \frac{1}{N} (| c_{00} \rangle \langle c_{00} | \otimes | \gamma^{0}_{00} \rangle \langle \gamma^{0}_{00} | + |d^{1}_{11}|^2 | c_{11} \rangle \langle c_{11} | \otimes | \gamma^{1}_{11} \rangle \langle \gamma^{1}_{11} |) . 
\end{equation}
The probability that the intermediate measurement yielded the eigenstate $c_1$ is
\begin{equation}
\text{Prob}[k = 1|\hat{\rho}(A,\Gamma ,B)] = \sum_{l,m} \text{Tr} (\hat{\rho}_{a, b} |\gamma^{1}_{lm} \rangle \langle \gamma^{1}_{lm} |)   =   \frac{|d^{1}_{11}|^2}{1 + |d^{1}_{11}|^2} .
\end{equation}
This is in agreement with Eq.~(26a) of Ref.~\cite{shimony1}.

As discussed in Sect.~III, one way to confirm that $\text{Prob}[k = 1]$ is time-symmetric is to interchange the pre- and post-selection states and consider time evolution backwards in time. Thus we begin with the state $| B \rangle $ in Eq.~(A6) and expressing that state in terms of the energy eigenstates, it follows (for $\epsilon = \pi/2g$) that the interaction between the quantum system and the IMS causes the evolution (in the backward-time direction)
\begin{equation}
\hat{U}(t_c - \epsilon , t_c)| B \rangle = \frac{1}{\sqrt{N}} [| c_{00} \rangle \otimes | \gamma \rangle + |d^{1}_{12}|^2 | c_{11} \rangle \otimes | \gamma \rangle + |d^{1}_{11}|^2 d^{1}_{12} |c_{12} \rangle |\gamma^{1}_{12} \rangle - d^{1}_{11} |d^{1}_{12}|^2 |c_{12} \rangle |\gamma^{1}_{12} \rangle ].
\end{equation}

The final step in the backward-time direction is projection onto the state $|a_i \rangle \otimes | \gamma \rangle $ given in Eq.~(A1) which leads to the original state $| A \rangle $ given in Eq.~(A1) after re-normalisation. Hence the same probabilities for the intermediate measurement of $\gamma^1_{11}$ are obtained when calculated in the reverse-time direction. As mentioned in Sect.~III, that conclusion differs from Ref.~\cite{shimony1} because in Ref.~\cite{shimony1} it is assumed that the state of the intermediate measuring apparatus is reset to $| \gamma \rangle$ at time $t_b$ so that the preselected state for the reverse-time evolution is $\frac{i}{\sqrt{N}} ( | c_{00} \rangle  + d^{1}_{11} | c_{12} \rangle ) \otimes | \gamma \rangle $ instead of the expression in Eq.~(A6).

\section{Consideration of the second counter-example of Ref.~\cite{shimony1}}

A second counter-example to time symmetry for a PPSE is given in Appendix A of Ref.~\cite{shimony1}. It addresses the suggestion that time symmetry would mean that the probabilities for a PPSE defined by an evolution of an initial state at time $t_a$ forward in time to a final state at $t_b$ should be the same when the initial and final states are interchanged (retaining the same forward evolution from $t_a$ to $t_b$). The counter-example shows that this criterion for time symmetry does not hold even for the simplest case. As discussed in Section III, the criterion of orthodox time-reversal which is closest to the above criterion would require that the probabilities were the same when the the initial and final states are interchanged {\it and} the initial and final states were time-reversed. The purpose of this Appendix is to confirm that this last criterion, corresponding to orthodox time-reversal symmetry, is satisfied by the counter-example. 

In the counter-example, the quantum system is in a four-dimensional Hilbert space with one non-degenerate eigenstate $| 0 \rangle $ with eigenvalue $k = 0$ and triply degenerate eigenstates $| 1i \rangle $, ($i = 1$, 2, 3)  with eigenvalue $k = 1$. For the original case, the pre- and post-selected states of the quantum system are
\begin{equation}
|a_i \rangle = \frac{1}{\sqrt{2}}(| 00 \rangle + | 11 \rangle )
\end{equation}
and
\begin{equation}
|b_j \rangle = \frac{1}{\sqrt{2}}(| 00 \rangle + | 12 \rangle ).
\end{equation}
It is assumed that the quantum system is subject to the time-evolution operator $\hat{U}(t_c,t_a) = \hat{U}(t_b,t_c) = \hat{U}$ both between $t_a$ and $t_c$ and between $t_c$ and $t_b$. In the $| 0 \rangle $, $| 1i \rangle $ basis, $\hat{U}$ is defined to be
\begin{equation}
\hat{U} = \left[
\begin{array}{cccc}
1 & 0 & 0 & 0 \\
0 & 0 & 0 & 1 \\
0 & 1 & 0 & 0 \\
0 & 0 & 1 & 0
\end{array}
\right].
\end{equation}
We first of all show that the expression for the state of a PPSE suggested above in Eq.~(8) leads to the same results as Ref.~\cite{shimony1}. The intermediate measurement simply distinguishes between the eigenvalues $k=0$ and $k=1$ so $\Gamma =  \{|\gamma^{0} \rangle \langle \gamma^{0} |, |\gamma^{1} \rangle \langle \gamma^{1} | \}$. The initial state at time $t_c$ is
\begin{equation}
|A;t_c \rangle = \frac{1}{\sqrt{2}}(| 00 \rangle \otimes | \gamma^0 \rangle + | 12 \rangle \otimes | \gamma^1 \rangle).
\end{equation}
After the final measurement interaction but prior to the final projection, using Eq.~(B3) the state is
\begin{equation}
|A;t_b \rangle = \frac{1}{\sqrt{2}}(\sum_i \langle b_i | 00 \rangle | b_i \rangle \otimes | \gamma^0 \rangle + \langle b_i | 13 \rangle | b_i \rangle \otimes | \gamma^1 \rangle).
\end{equation}
Projecting onto $| b_j \rangle $ given in Eq.~(B2), the final state of the PPSE is 
\begin{equation}
|B;t_b \rangle = \frac{1}{\sqrt{2}}(| 00 \rangle  + | 12 \rangle ) \otimes | \gamma^0 \rangle \otimes | \beta_j \rangle \otimes | \alpha_i \rangle)
\end{equation}
and using $\hat{U}^{\dag}$ from Eq.~(B3), 
\begin{equation}
|B;t_c \rangle = \frac{1}{\sqrt{2}}(| 00 \rangle  + | 11 \rangle ) \otimes | \gamma^0 \rangle \otimes | \beta_j \rangle \otimes | \alpha_i \rangle).
\end{equation}

Using Eqs.~(B4) and (B7) in Eq.~(10), Prob$[k=1|A;\{\gamma^k\},B]=1$ in agreement with Eq.~(A6) of Ref.~\cite{shimony1}.

If the initial and final state in Eqs.~(B1) and (B2) are interchange so that $|a_i \rangle = (| 00 \rangle + | 12 \rangle )/ \sqrt{2}$ and $|b_j \rangle = (| 00 \rangle + | 11 \rangle )/ \sqrt{2}$, one finds
\begin{equation}
|A;t_c \rangle = \frac{1}{\sqrt{2}}(| 00 \rangle \otimes | \gamma^0 \rangle + | 13 \rangle \otimes | \gamma^1 \rangle).
\end{equation}
and
\begin{equation}
|B;t_c \rangle = \frac{1}{2}(| 00 \rangle  + | 13 \rangle ) \otimes (| \gamma^0 \rangle + | \gamma^1 \rangle ) \otimes | \beta_j \rangle \otimes | \alpha_i \rangle).
\end{equation}
so that from Eq.~(10), Prob$[k=1|A;\{\gamma^k\},B]=1/2$ in agreement with Eq.~(A8) of Ref.~\cite{shimony1}.

As mentioned, the orthodox view is that time symmetry should tested by both the interchanging {\it and} time-reversing the states.  Assuming the Hamiltonian in the example is time symmetric, the time-reversal operator $\hat{\Theta}$ can be found from the condition $\hat{U} \hat{\Theta} = \hat{\Theta}\hat{U}^{\dag}$ and in the same the $| 0 \rangle $, $| 1i \rangle $ basis 
\begin{equation}
\hat{\Theta} = \frac{1}{\sqrt{3}} \left[
\begin{array}{cccc}
\sqrt{3} & 0 & 0 & 0 \\
0 & 1 & r & 1 \\
0 & r & 1 & 1 \\
0 & 1 & 1 & r
\end{array}
\right]K
\end{equation}
where $r = -(1 + i \sqrt{3})/2$ and $K$ is the operator which takes the complex-conjugate of complex numbers.

Therefore to consider the time-reversal of the original PPSE specified by $| a_i \rangle $ and $| b_j \rangle $ given in Eqs.~(B1) and (B2), the new initial state of the quantum system is
\begin{equation}
| \tilde{a}' \rangle = \hat{\Theta } | \tilde{b} \rangle = \frac{1}{\sqrt{6}} [\sqrt{3} | 00 \rangle - \frac{1}{2}(1 + i \sqrt{3})| 11 \rangle + | 12 \rangle + | 13 \rangle]
\end{equation}
and the new final state of the quantum system is
\begin{equation}
| \tilde{b}' \rangle = \hat{\Theta } | \tilde{b} \rangle = \frac{1}{\sqrt{6}} [\sqrt{3} | 00 \rangle + | 11 \rangle - \frac{1}{2}(1 + i \sqrt{3})| 12 \rangle + | 13 \rangle] .
\end{equation}
One then finds that
\begin{equation}
|A;t_c \rangle = \frac{1}{\sqrt{6}}[\sqrt{3} | 00 \rangle \otimes | \gamma^0 \rangle + (| 12 \rangle - \frac{1}{2}(1 + i \sqrt{3})| 13 \rangle + | 11 \rangle ) \otimes | \gamma^1 \rangle ] \otimes | \beta \rangle \otimes | \alpha_i \rangle)
\end{equation}
and
\begin{equation}
|B;t_c \rangle = \frac{1}{\sqrt{6}}[\sqrt{3} | 00 \rangle \otimes | \gamma^0 \rangle + | 11 \rangle  \otimes | \gamma^1 \rangle ] \otimes | \beta_j \rangle \otimes | \alpha_i \rangle).
\end{equation}

Using Eqs.~(B13) and (B14) in Eq.~(10), Prob$[k=1|A;\{\gamma^k\},B]=1$ which shows that the PPSE of the counter-example is time-symmetric on the orthodox view of time symmetry. It differs from Ref.~\cite{shimony1} because the time reversal of the states was omitted in Ref.~\cite{shimony1}.

\end{document}